\documentstyle[12pt,epsfig]{article} 
\begin{document}

\baselineskip=16pt  

\hyphenation{
mani-fold
mani-folds
}


\def\Bbb{\bf}

\def\BbbR{{\Bbb R}}
\def\BbbZ{{\Bbb Z}}
\def\BbbC{{\Bbb C}}

\def\casehalf{{\case{1}{2}}}

\def\morsef{{f}}

\def\so{{\hbox{${\rm SO}(2,1)$}}}

\def\eps{\epsilon}

\def\Re{{\rm Re}}
\def\Im{{\rm Im}}



\title{Bianchi I  Quantum cosmology in the Bergmann-Wagoner theory}
\author{Luis O. Pimentel
\thanks{email: lopr@xanum.uam.mx.
} \\
Departmentamento de F\'{\i}sica,\\
Universidad Aut\'onoma Metropolitana, Iztapalapa
\\
P.O. Box 55-534,\\
CP 09340, M\'exico D. F., MEXICO
}
\date{}
\maketitle
\begin{abstract}%
\centerline{\it Dedicated to Heinz Dehnen in honour of his $65^{th}$
birthday.}
\bigskip

The Wheeler-DeWitt equation is considered in the context of generalized
scalar-tensor theories of gravitation for the Bianchi type I cosmology.
Exact solutions are found for two selfinteracting potentials and arbitrary
coupling function. The WKB wavefunctions are obtained and a family of
solutions satisfying the Hawking-Page regularity conditions of wormholes
are found.
\end{abstract}

\newpage

\section{Introduction}

In this paper we consider quantum cosmological models for Bianchi type I 
 models in the generalized vacuum scalar--tensor theories of gravity.
Interest in these theories has been widespread in recent years in
connection with inflation and string theories. They are defined by the
action

\begin{equation}
\label{action}
A=\int d^4 x \sqrt{-g} e^{-\Phi}
 \left[  R - \omega (\Phi ) \left(
\nabla \Phi \right)^2 -2\lambda (\Phi) \right] ,
\end{equation}
where $R$ is the Ricci curvature of the space--time and
$g$ is the determinant of the
metric $g_{\mu\nu}$ \cite{ST}. The dilaton field $\Phi$
plays the role of a time--varying gravitational constant
and may self--interact through a potential  $\lambda (\Phi )$
( the usual Brans-Dicke scalar field is $\phi=exp(-\Phi)$ ).
The function $\omega (\Phi)$ plays the role of a coupling  function 
between the dilaton and graviton.  Each scalar--tensor theory is
characterized 
by specific  functional forms of
$\omega (\Phi)$ and $\lambda (\Phi)$. A cosmological constant in the
gravitational sector of the theory corresponds to
the special case where $\lambda (\Phi)$ is a space--time constant.
 Among others we have the case of Brans--Dicke theory, where $\omega
(\Phi)$ is a 
space--time constant and $\lambda (\Phi )$ is absent
\cite{BD}.
It is known that inflationary solutions exist  in a wide class of
scalar--tensor cosmologies \cite{PSC}, therefore these theories are
relevant to the 
study of  the very early Universe \cite{INF}.
Also, higher--order \cite{W} and higher--dimensional \cite{HD}
theories of gravity may be expressed in a scalar--tensor
form after suitable field redefinitions. Brans--Dicke theory
with $\omega =-1$ corresponds to a truncated version
of the string effective action \cite{STRING}.

Point symmetries associated with action (\ref{action})
have been discussed previously within the
context of the spatially isotropic Friedmann Universes \cite{PS,PS2} and
Bianchi models \cite{L}. It was found that $\omega (\Phi)$ and $\lambda
(\Phi)$ must be related in a  certain way if the field equations 
are to be symmetric.

In this paper we consider the Wheeler-DeWitt equation (WDW) for the simplest
anisotropic model, namely, the Bianchi type I cosmology. 
Some time ago the canonical formulation 
of the Brans-Dicke theory was considered by Toton and by Matzner et al.
\cite{TM}. For a recent review of WDW in different theories of gravitation see Ref. \cite{amac}
The line element for the class of spatially homogeneous space-times is
given by
\begin{equation}
\label{metric}
ds^2=-dt^2 +h_{ab} \omega^a \omega^b, \qquad a,b=1,2,3 ,
\end{equation}
where $h_{ab}(t)$ is a function of cosmic time $t$ and
represents the metric on the surfaces of homogeneity and
$\omega^a$ are one--forms. These models have
a topology $R\times G_3$, where $G_3$ represents a Lie group of isometries
that acts transitively on the space--like three--dimensional orbits
\cite{RS}. The Lie algebra of $G_3$ admits
the structure  constants ${C^a}_{bc}=m^{ad}\epsilon_{dbc}+
{\delta^a}_{[b} a_{c]}$, where $m^{ab}$ is a symmetric matrix, $a_c
\equiv {C^a}_{ac}$ and $\epsilon_{abc} = \epsilon_{[abc]}$.
The Jacobi identity
${C^a}_{b[c} {C^b}_{de]} =0$ is only satisfied if $m^{ab}a_b =0$, so
$m^{ab}$ must be transverse to $a_b$ \cite{10}. The model belongs to the
Bianchi class A if $a_b=0$ and to the class B if $a_b \ne 0$.
A basis may be found such that $a_b =(a,0,0)$ and
$m^{ab} ={\rm diag} \left[ m_{11},m_{22},m_{33} \right]$,
where $m_{ii}$ take the values  $\pm 1$ or $0$.
In the Bianchi class A,
the Lie algebra is uniquely determined up to isomorphisms
by the rank and signature of $m^{ab}$. The six possibilities are
$(0,0,0)$, $(1,0,0)$, $(1,-1,0)$, $(1,1,0)$, $(1,1,-1)$ and $(1,1,1)$ and
these correspond, respectively, to the Bianchi types I, II, ${\rm VI}_0$,
${\rm VII}_0$, VIII and IX. Finally,
the three-metric may be parametrized by $h_{ab}(t)=e^{2\alpha (t)}
\left( e^{2\beta (t)} \right)_{ab}$, where $e^{3\alpha}$ represents the
effective spatial volume of the Universe and
\begin{equation}
\beta_{ab} \equiv
{\rm diag} \left[ \beta_+ +\sqrt{3}\beta_-,\beta_+ -\sqrt{3}\beta_- , -2
\beta_+ \right]
\end{equation}
is a traceless matrix that determines the anisotropy in the models.

The configuration space $Q$ for the Bianchi
models derived from action (\ref{action}) is
therefore four--dimensional and is
spanned by  $\{ q_n \equiv \alpha , \Phi ,\beta_{\pm} \}$.
The Lagrangiandensity $L (q_n ,\dot{q}_n )$ is
defined by $A=\int dt L (q_n , \dot{q}_n )$,
where a dot denotes differentiation
with respect to cosmic time. It may be derived by substituting
the metric (\ref{metric}) into the action (\ref{action}) and
integrating over the spatial variables. This procedure is unambiguous
for the class A cosmologies and the action for these models simplifies to
\cite{L}
\begin{equation}
\label{actionbianchi}
A = \int dt e^{3\alpha -\Phi} \left[
6\dot{\alpha} \dot{\Phi} -6\dot{\alpha}^2 +6\dot{\beta}^2_+ +
6\dot{\beta}^2_- +\omega (\Phi)
\dot{\Phi}^2  -2\lambda (\Phi ) +e^{-2\alpha} U(\beta_{\pm})
\right]    ,
\end{equation}
where
\begin{equation}
\label{potentialA}
U(\beta_{\pm}) = -e^{-4 \alpha} \left( m_{ab} m^{ab} -\frac{1}{2} m^2
\right)
\end{equation}
is the curvature potential,
$m \equiv {m^a}_a$ and indices are raised and lowered with $h^{ab}$
and $h_{ab}$, respectively \cite{WALD}. In the case of the
type B models, a divergence may arise because the three--curvature
contains a term proportional to $a_b a^b$  \cite{Mac}. In view of this
difficulty,
we do not consider these models further.

To proceed we take the following changes of variables,

\begin{equation}
\label{changev}
dt=e^{\Phi /2}d\tau , \; x={\alpha-\Phi
/2},\;y=\int{\sqrt{\frac{3+2\omega(\Phi)}{12}}}d\Phi,
\end{equation}
and the action is now

\begin{equation}
\label{action1}
A = \int d\tau  \left[6e^{3x}
\left\{y'^2- {x'^2} +{\beta '}^2_+ +{\beta '}^2_- -\Lambda (y )\right\} 
+ e^{x} U(\beta_{\pm})
\right]    ,
\end{equation}
where 
\begin{equation}
\label{}
\Lambda(y):=e^{\Phi}\lambda(\Phi)/3.
\end{equation}
The above action means that the Lagrangian is

\begin{equation}
L=\left[6e^{3x}\left\{y'^2- {x'^2} +{\beta '}^2_+ +
{\beta '}^2_-  -\Lambda (y )\right\} + e^{x} U(\beta_{\pm})
\right].
\end{equation}
From here we can calculate the canonical momenta and the Hamiltonian

\begin{eqnarray}
\label{hamiltonian}
\pi_x=-12e^{3x}x',\;\pi_y=12e^{3x}y',\;\pi_+=12e^{3x}\beta_+',
\;\pi_-=12e^{3x}\beta_-',\\
H=\frac{e^{-3x}}{24}[-{\pi_x^2}+{\pi_y^2}
+{\pi_+^2}
+{\pi_-^2}+144e^{6x}{\Lambda (y )} -24 e^{4x} {U(\beta_{\pm})}].
\end{eqnarray}
Then the Wheeler-DeWitt equation 
 follows from the  canonical quantization of 
$H=0$, i.e., the canonical momenta in Eq.(\ref{hamiltonian}) are converted
into operators in the standard way, in general, owing to the ordering 
ambiguity we have

\begin{equation}
\pi_x^2 \to -\frac{1}{x^B}\frac{\partial}{\partial x}
\left(x^B\frac{\partial}{\partial x}\right) ,
\nonumber
\end{equation}
therefore the WDW equation $H\Psi(x,y)=0$ for an
arbitrary factor ordering, encoded in the $B$ parameter,
is

\begin{equation}
\left[\partial^2_x -B\partial_x - \partial^2_y - \partial^2_+  
- \partial^2_-  
- 24 e^{4x} U(\beta_{\pm})
+ 144 e^{6x}\Lambda (y)\right]\Psi(x,y,\beta_{\pm})=0.
\label{wdw}
\end{equation}
In the following section we try to solve the Wheeler-DeWitt equation by
separation of variables in the simplest of the homogeneous cosmologies,
namely the Bianchi Type I case in which the potential $ U(\beta_{\pm})$
vanishes.

\section{Exact solutions for Bianchi I}

We now consider the case of Bianchi type I cosmological model for which
the potential 
$ U(\beta_{\pm})$ vanishes identically. Furthermore we restrict ourselves
to the case when
$\Lambda (y)=\Lambda_o/144={\rm constant} $,i.e. $\lambda (\phi)\propto
\phi$; this form of the potential has been used previously to obtain
classical solutions in Bianchi type I vacuum cosmology \cite{BDB} and for
isotropic models with a barotropic fluid\cite{GS,LM}; this potential is
one of those that could produce inflation\cite{PSC}. The choices that we
have made simplify the  WDW equation and allows us to obtain exact
solutions by means of separation of variables,

\begin{equation}
\Psi(x,y,\beta_{\pm})=X(x)Y(y)F(\beta_{+})G(\beta_{-}).
\end{equation}
This implies that equation(\ref{wdw} )gives the separeted equations

\begin{equation}
X''-BX'+[\kappa_0^2+\Lambda_0 e^{6x}]X=0,
\end{equation}

\begin{equation}
Y''+k^2Y=0,
\end{equation}

\begin{equation}
F''+k_+^2F=0,
\end{equation}

\begin{equation}
G''+k_-^2G=0,
\end{equation}
where 
\begin{equation}
\kappa_0^2= k_+^2+k_-^2+
k^2.
\end{equation}
The solutions to these equations give the wavefunction:

\begin{eqnarray}
\Psi(x,y,\beta_{\pm})&=&e^{Bx/2}
[c_1
J_{\nu}(\frac{\sqrt{\Lambda_0}}{3}e^{3x})+c_2Y_{\nu}(\frac{\sqrt{\Lambda_0}}{3}e^{3x})]
[c_3e^{iky}+c_4e^{-iky}]\nonumber \\
&&[c_5e^{ik_+\beta_+}+c_6e^{-ik_+\beta_+}][c_7e^{ik_-\beta_-}+c_8e^{ik_-\beta_-}],
\end{eqnarray}
where
\begin{equation}
\nu=\frac{\sqrt{-B-4\kappa_0^2}}{6},
\end{equation}
and the $c_i$ are constants.
By superposition of these solutions, wavefunctions satisfying different
boundary conditions can be obtained. In the following subsection we
consider the case of wormholes.

\subsection{Wormhole solution}

Quantum wormholes can be regard as special class of solutions to 
the Wheeler-DeWitt equation with certain boundary conditions\cite{7}: 
$i)$ the wavefunction is exponentially damped for large spatial geometry, 
i.e., when $\alpha \to \infty$, 
$ii)$ the wavefunction is regular when the spatial geometry degenerates, 
i.e., the wavefunction does not oscillate when $\alpha \to -\infty$. These
conditions 
are known as the Hawking-Page regularity conditions (HP). In what follows 
we show, with a particular factor ordering, the 
explicit form of  a wavefunction that satisfies the HP regularity 
conditions. 

The following wavefunction can be obtained by superposition of the
solutions of the previous section or it can be substituted into the WDW
equations to check that it is a particular solution

\begin{equation}
\Psi(x,y,\beta_{\pm})={e}^{m {Cosh}[ny+p\beta_{+}q\beta_{-}+r]exp(3x)},
\end{equation}
where r is an arbitrary real parameter, the factor ordering $B$ and the
constants $m,n,p,q$ satisfy the relations

\begin{equation}
B=3+\Lambda_0/3,n^2+p^2+q^2=9, -(n^2+p^2+q^2)m^2=\Lambda_0.
\end{equation}
From these relations we see that for a wormhole we require a negative
$\Lambda_0$. If from the last relation we take the negative root for $m$,
it is easy to check that  the wavefunction is exponentially damped for
large spatial geometry, i.e., when $\alpha \to \infty$ ($x \to
\infty$) and also that the wavefunction does not oscillate when $\alpha
\to -\infty$( $x \to -\infty$).

\subsection{WKB solution}

Here we want to obtain the WKB wavefunction in the form 

\begin{equation}
\Psi(\eta,\xi,\beta_{\pm})=e^{i[S(x,y,\beta_+,\beta_-)]}.
\end{equation}
After substitution into the WDW equation the Hamilton-Jacobi equation
results and separating variables it is straightforward to obtain the
following solution,

\begin{equation}
S=p_yy+p_+\beta_++p_-\beta_- \pm \left [ {\frac{{\sqrt{{k^2} +
\Lambda_0\,{e^{6\,x}}}}}{3}} - 
  {\frac{k\,{\rm ArcTanh}({\frac{{\sqrt{{k^2} + 
             \Lambda_0\, {e^{6\,x}}}}}{k}})}{3}}-k\, x\right ],
\end{equation}
where $k^2=p_y^2+p_+^2+ p_-^2$. Once that we have solved the
Hamilton-Jacobi equation it is possible to find the  classical solution,
we do not do that here because they were obtained by Banerjee et
al.\cite{BDB}, solving the field equations.

\section{Another potential}

We consider now another potential function for which it is possible to
obtain exact solutions to WDW equation. In Eq.(\ref{action1}) we change
the time $\tau$ in the following way $d\tau =e^{-x}d\sigma$, the action
becomes,

\begin{equation}
\label{action2}
A = 6\int d\sigma  \left[e^{4x}
\left\{y'^2- {x'^2} +{\beta '}^2_+ +{\beta '}^2_- \right\} -e^{2x}
\Lambda (y )
\right]    ,
\end{equation}
here the prime means derivative with repect to $\sigma$.
We introduce a change of variables and an explicit potential,

\begin{equation}
\eta=e^{2x}{\rm Cosh} (2y),\, \xi=e^{2x}{\rm Sinh
}(2y),\,\Lambda(y)=\Lambda_1{\rm Cosh}(2y)+ \Lambda_2{\rm Sinh }(2y).
\end{equation}
In the new time and variables, the action, Lagrangean, Hamiltonian can be
calculated with the following results,

\begin{equation}
\label{action3}
A = 6\int d\sigma  \left[\frac{1}{4}
(\xi '^2- \eta '^2) +(\eta^2-\xi^2)({\beta_+ '}^2 +{\beta_- '}^2 )
-\Lambda_1 \eta-\Lambda_2 \xi
\right]    ,
\end{equation}

\begin{equation}
L=6[\frac{1}{4}(\xi '^2- \eta '^2) +(\eta^2-\xi^2)({\beta_+ '}^2 +{\beta_-
'}^2 ) -\Lambda_1 \eta-\Lambda_2 \xi ], 
\end{equation}

\begin{equation}
H=\frac{1}{24}[\pi_\xi^2-\pi_\eta^2+(\eta^2-\xi^2)({\pi_+}^2 +{\pi_-}^2
)+24\Lambda_1 \eta +24\Lambda_2 \xi ]. 
\end{equation}
The WDW equations in this case is

\begin{equation}
\left[\partial^2_\xi - \partial^2_\eta  + (\eta^2-\xi^2)(\partial^2_+  
+ \partial^2_-)  
-  24\Lambda_1 \eta -24\Lambda_2 \xi 
\right]\Psi(\eta,\xi,\beta_{\pm})=0.
\label{wdw2}
\end{equation}
Assuming separation of variables,

\begin{equation}
\Psi(\eta,\xi,\beta_{\pm})=E(\eta)X(\xi)P(\beta_+)M(\beta_-),
\end{equation}
the corresponding equations are
\begin{equation}
E''+ [k_{\perp}^2 \eta^2+24\Lambda_1\eta+k^2]E=0,
\end{equation}

\begin{equation}
X''+ [k_{\perp}^2 \xi^2-24\Lambda_2\xi+k^2]X=0,
\end{equation}

\begin{equation}
P''+ k_+^2P=0,
\end{equation}

\begin{equation}
M''+ k_-^2M=0,
\end{equation}
where $k_+,k_-$ and $k$ are arbitrary separation constants and
$k_{\perp}^2=k_+^2+k_-^2$.
The solutions are 

\begin{eqnarray}
E&=&c_1{ \;\; {}_1F_1}
(a_1,1/2,z_1)+c_2z_1^{1/2}{ \;\; {}_1F_1}
(a_1+1/2,3/2,z_1),\nonumber \\
a_1&=&\frac{1}{4}[{-k^2+(\frac{12\Lambda_1}{k_\perp})^2}{\sqrt{-k_\perp
^2}}+1],\,z_1=\sqrt{-k_\perp^2}(\eta+\frac{12\Lambda_1}{k_\perp^2})^2,
\end{eqnarray}

\begin{eqnarray}
X&=&c_3{ \;\; {}_1F_1}
(a_2,1/2,z_2)+c_4z_2^{1/2}{ \;\;{}_1F_1}
(a_2+1/2,3/2,z_2),\nonumber \\
a_2&=&\frac{1}{4}[{-k^2+(\frac{12\Lambda_2}{k_\perp})^2}{\sqrt{-k_\perp
^2}}+1],\,z_2=\sqrt{-k_\perp^2}(\xi+\frac{12\Lambda_2}{k_\perp^2})^2,
\end{eqnarray}

\begin{equation}
P=c_5e^{ik_{+} {\beta_{+}}}+c_6e^{-ik_{+}{\beta_{+}}},
\end{equation}

\begin{equation}
M=c_7e^{ik_{-}{\beta_{-}}}+c_8e^{-ik_{-}{\beta_{-}}},
\end{equation}
where ${ \;\; {}_1F_1}
$ is the confluent hypergeometric function. More solutions can be obtained
by superposition.

\subsection{WKB solution}

Again for the new potential we want to find the WKB wavefunction,
\begin{equation}
\Psi(\eta,\xi,\beta_{\pm})=e^{i[V(\eta)+W(\xi)+ (p_+\beta_ +) + (p_-\beta_-)]}.
\end{equation}
After substitution into the WDW equation we have

\begin{equation}
V_\eta^2-p_\perp^2\eta^2-24 \Lambda_1\eta-p^2=0,
\end{equation}

\begin{equation}
W_\xi^2-p_\perp^2\xi^2+24 \Lambda_2\xi-p^2=0,
\end{equation}
where $p_\perp^2=p_+^2 + p_-^2$. The solutions are

\begin{eqnarray}
V&=&\left( {\frac{6\,{\Lambda_1}}{{{{p_\perp}}^2}}} + 
     {\frac{\eta}{2}} \right) \,
   {\sqrt{{p^2} + 24\,{\Lambda_1}\,\eta + 
       {{{p_\perp}}^2}\,{\eta^2}}} + \nonumber \\
& &  {\frac{\left( -144\,{{{\Lambda_1}}^2} + 
        {p^2}\,{{{p_\perp}}^2} \right) \,
      \log ({\frac{2\,\left( 12\,{\Lambda_1} + 
              {{{p_\perp}}^2}\,\eta \right) }{{p_\perp
            }}} + 2\,{\sqrt{{p^2} + 
             24\,{\Lambda_1}\,\eta + 
             {{{p_\perp}}^2}\,{\eta^2}}})}{2\,
      {{{p_\perp}}^3}}},
\end{eqnarray}

\begin{eqnarray}
W&=&\left( -{\frac{6\,{\Lambda_2}}{{{{p_\perp}}^2}}} + 
     {\frac{\eta}{2}} \right) \,
   {\sqrt{{p^2} -24\,{\Lambda_2}\,\eta + 
       {{{p_\perp}}^2}\,{\eta^2}}} + \nonumber \\
& &  {\frac{\left( -144\,{{{\Lambda_2}}^2} + 
        {p^2}\,{{{p_\perp}}^2} \right) \,
      \log ({\frac{2\,\left( -12\,{\Lambda_2} + 
              {{{p_\perp}}^2}\,\eta \right) }{{p_\perp
            }}} + 2\,{\sqrt{{p^2} - 
             24\,{\Lambda_2}\,\eta + 
             {{{p_\perp}}^2}\,{\eta^2}}})}{2\,
      {{{p_\perp}}^3}}}.
\end{eqnarray}

\section{Final remarks}

In this work we have obtained exact solutions for the WDW equation using
the general scalar tensor theory of gravitation with arbitary coupling
function $\omega (\Phi)$ and two specific selfinteracting potentials
$\lambda(\Phi)$. The WKB wavefunctions were obatained for both cases.  The
solutions here obtained are for arbitrary large anisotropies. We hope that
these solutions should prove to be useful in the study of the issues of
quantum cosmology. In the past quantum cosmology using general relativity
in homogeneous spacetimes with small anisotropies  and a scalar field were
considered by  Lukash and Schmidt \cite {LS}, and Amsterdamski \cite{AM}.
Lidsey \cite{LHA} has considered the wavefunctions in a highly anisotropic
cosmologies with a massless minimally coupled scalar field.  More recently
Bachmann and Schmidt \cite {BS} have considered the case of arbitrary
anisotropies in Bianchi I quantum cosmology.

\end{document}